\documentstyle[11pt,paspconf,epsf]{article}

\begin{document}

\title{ The Low-Mass Stellar IMF at High Redshift:
 Faint Stars in 
 the Ursa Minor Dwarf Spheroidal Galaxy{\footnote{}}}

\author{Rosemary F.G.~Wyse
\altaffilmark{2}, 
Gerard Gilmore\altaffilmark{3}, 
Sofia Feltzing\altaffilmark{4} \&  Mark Houdashelt\altaffilmark{2}}
\altaffiltext{1}{Based on observations with the NASA/ESA Hubble Space Telescope, obtained at STScI, operated by AURA Inc, under NASA contract NAS5-26555}
\altaffiltext{2}{Physics \& Astronomy Department, Johns Hopkins University, Baltimore, MD 21218}
\altaffiltext{3}{Institute of  Astronomy, Madingley Road, Cambridge CB3 0HA, UK}
\altaffiltext{4}{Lund Observatory, Box 43, 221 00 Lund, Sweden}





\keywords{stars: low-mass, brown dwarfs; stars: luminosity function,
mass function; galaxies: dwarf; (cosmology:) dark matter}

\section{Introduction}

Low-mass stars, those with main-sequence lifetimes that are of order
the age of the Universe, provide unique constraints on the Initial
Mass Function (IMF) when they formed.  Star counts in systems with simple
star-formation histories are particularly straightforward to
interpret, and those in `old' systems allow one to determine the
low-mass stellar IMF at large look-back times and thus at high
redshift.  We present the faint stellar luminosity function in an
external galaxy, the Ursa Minor dwarf Spheroidal (dSph).  This
relatively-nearby (distance $\sim 70$kpc) companion galaxy to the
Milky Way has a stellar population with narrow distributions of age
and of metallicity (e.g. Hernandez, Gilmore \& Valls-Gabaud 1999), 
remarkably similar to that of a classical halo globular
cluster such as M92 or M15, i.e. old and metal-poor ([Fe/H] $ \sim
-2.2$~dex).  The integrated luminosity of the Ursa Minor dSph ($L_V
\sim 3 \times 10^5 L_\odot$) is also similar to that of a globular
cluster. However, the central surface brightness of the Ursa Minor
dSph is only 25.5 V-mag/sq arcsec, corresponding to a central
luminosity density of only 0.006$L_\odot$pc$^{-3}$, many orders of
magnitude lower than that of a typical globular cluster.  Further,
again in contrast to globular clusters, its internal dynamics are
dominated by dark matter, with $(M/L)_V \sim 80$, based on the
relatively high value of its internal stellar velocity dispersion
(Hargreaves {\it et al.} 1994; see review of Mateo 1998).
Faint star counts in the Ursa Minor dSph thus allow determination of the 
low-mass IMF in a dark-matter-dominated external galaxy, in 
which the stars formed at high redshift.

\section{Observations}

We obtained deep imaging data with the Hubble Space Telescope, using
WFPC2 (V-606 \& I-814), STIS (LP optical filter) and NICMOS
(H-band), in a field close to the center of the Ursa Minor dSph
(program GO~7419: PI~Wyse, Co-Is Gilmore, Tanvir, Gallagher \&
Smecker-Hane; due to successive failures of HST the data acquisition
phase of this project remains ongoing).  The estimation of the 
contamination by foreground stars and background
galaxies required acquisition of similarly-exposed data for an offset
field $\sim 2$ tidal radii away from the Ursa Minor dSph,  at
similar Galactic coordinates to the UMi field ($\ell = 105^o, \,
b=45^o$); this field shows  no evidence for Ursa Minor member stars.

\subsection  { WFPC2 Data for the Ursa Minor dSph}

The WFPC2 data are discussed in more detail in Feltzing {\it et al.}
(1999). They consist of $8 \times 1200$s in each of F606W and F814W
filters. Standard HST data reduction techniques were followed using
the IRAF STSDAS routines, with photometry on the reduced images using
DAOPHOT and TinyTim psf's.  The scatter in the zero points and
photometric calibrations is $\sim 6\%$, 
which is small compared to the 0.5~magnitude binning we adopt for
the luminosity functions below.  Identical procedures were
applied to both the UMi and the offset-field datasets.  The completeness of
the data was determined by adding artificial stars to the original
images and then re-processing them.  The luminosity
functions discussed below include only stars 
detected in both V and I.

\subsection  { STIS LP data for the Ursa Minor dSph \& M15}

We obtained seven 2900s exposures of each of the Ursa Minor field and
the offset field.  We also obtained a single
exposure (1200s) of the globular cluster M15 in a field 
with extant deep WFPC2 V and I data (Piotto, Cool \& King 1997).  
The metallicity and age of 
the stars in  M15 are very similar to those in the Ursa Minor dSph and 
the STIS LP magnitudes for the Ursa Minor dSph 
may be transformed to I-814 using the extant M15 I-814 data (courtesy 
of G.~Piotto). 
Standard data reduction procedures as above were applied to our 
data.

\section{Faint Luminosity Functions}

\subsection {Globular Clusters} 

We utilise comparisons between the data for the Ursa Minor dSph and
globular clusters of the same metallicity and age (M92 and M15).  A
direct comparison between the luminosity functions corresponds to a
comparison between the underlying stellar mass functions.
However, the
stellar mass function in globular clusters may be modified by both
internal and external effects and thus in not all cases is the
present-day mass function a good estimate of the initial mass
function.  The Piotto, Cool \& King (1997) data were obtained at
intermediate radius within M15 and M92, where the effects of mass
segregation are expected to be small and the local faint luminosity
function should be a good estimate of the global faint luminosity
function.  Further, as discussed by Piotto \& Zoccali (1999), M92 and
M15 have fairly steep main-sequence luminosity functions, while other
globular clusters, particularly those for which external dynamical effects such as tidal
shocking by the disk or bulge of the Milky Way may be important, have
flatter faint luminosity functions (these differences occur fainter 
than the limits of the Ursa Minor dSph data). Thus we will assume that the present-day faint luminosity
functions of M92 and M15 may be taken as reasonable estimates of the
initial functions.

\subsection{ Ursa Minor  Faint Luminosity Function}

The derived, completeness-corrected, Color-Magnitude Diagram-based, 
WFPC2 V-band and I-band 
luminosity
functions for the Ursa Minor dSph are compared in  Figure~1 to those
for the globular cluster M92 (courtesy of 
G.~Piotto). 
We find  a remarkable similarity between
them, down to our 50\% completeness limit, which
corresponds to $\sim 0.4M_\odot$ (Baraffe {\it et al.} 1997 models).

\begin{figure}
\vskip 1cm
\plotfiddle{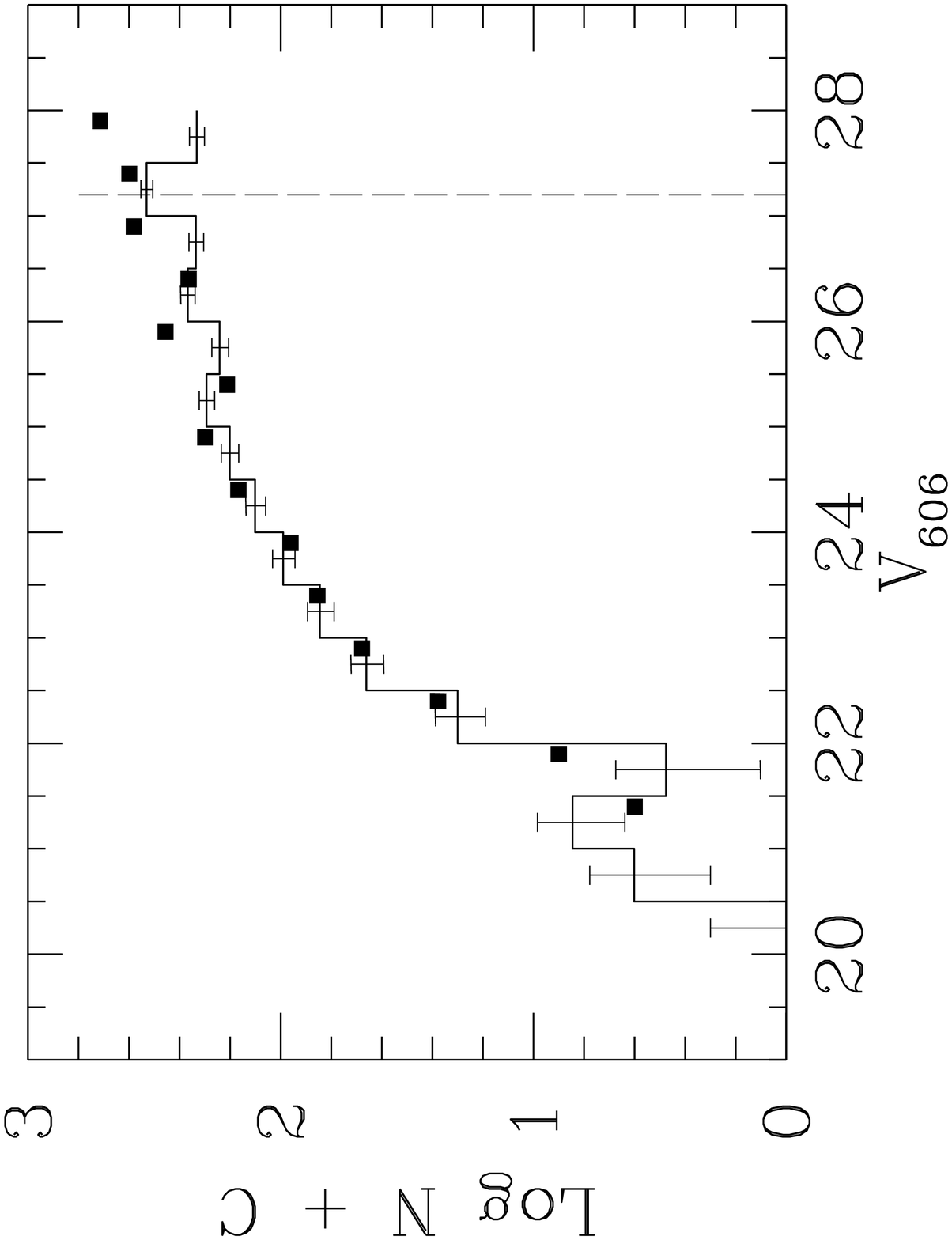}{4cm}{270}{25}{25}{-190}{150}
\vskip -4.5cm
\plotfiddle{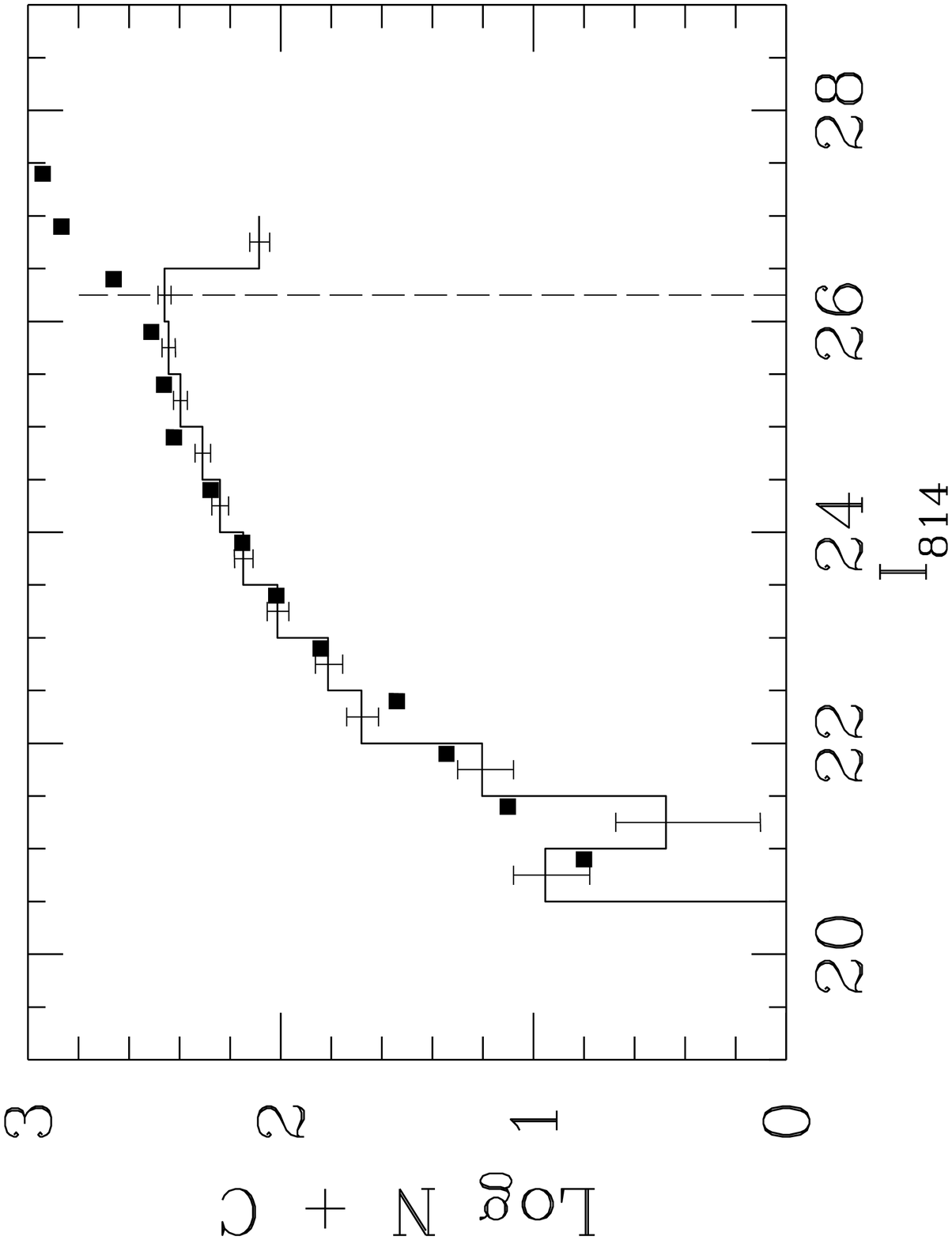}{4cm}{270}{25}{25}{0}{150}
\vskip -0.25truecm
\caption{Comparison between the completeness-corrected Ursa Minor
luminosity functions (histograms; 
50\% completeness indicated by the vertical dashed line) 
in the V-band (left panel) and the I-band (right panel)  and the same 
for M92 (points; from Piotto {\it et al.} 1997).}
\end{figure}

A direct comparison between our STIS optical luminosity functions for
Ursa Minor and  for the globular cluster M15 is shown in Figure 2 (left panel; only those
bins at least 50\% complete are shown).  Again,
there is very good agreement in slopes of luminosity functions, and
again our 50\% completeness limit for the Ursa Minor data 
is around $0.4M_\odot$.

\begin{figure}
\vskip 1cm
\plotfiddle{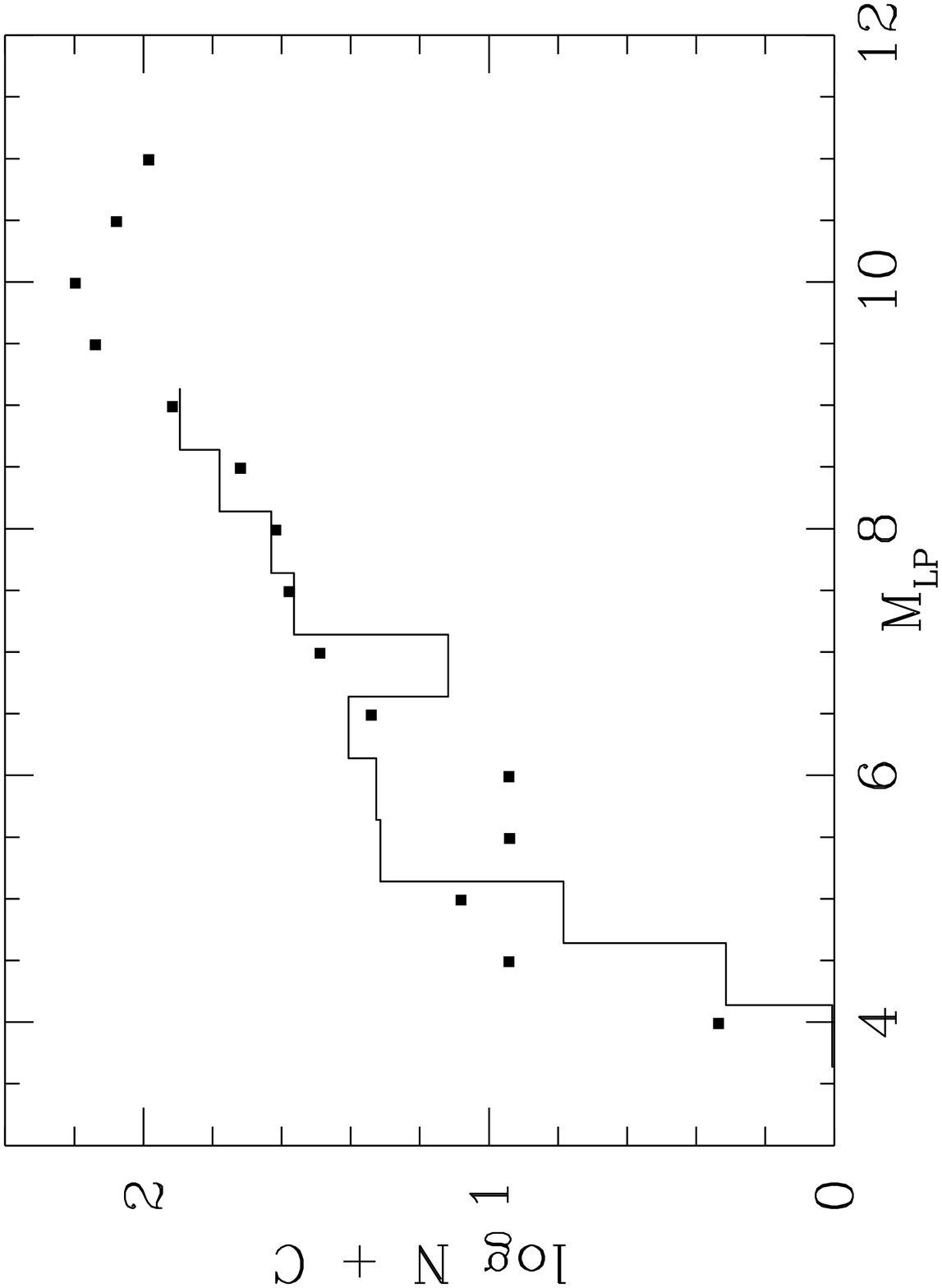}{4cm}{270}{25}{25}{-190}{150}
\vskip -4.5cm
\plotfiddle{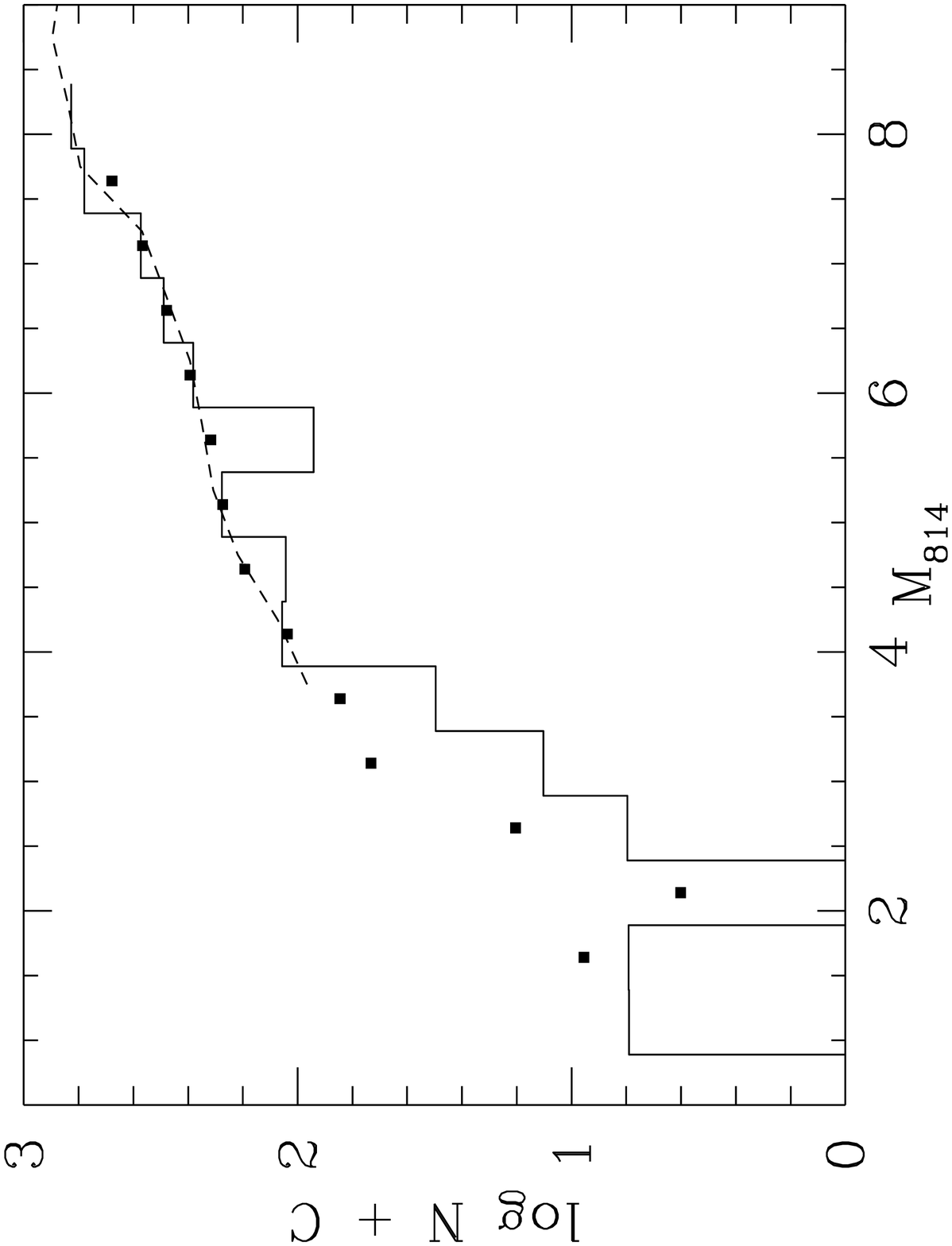}{4cm}{270}{25}{25}{0}{150}
\vskip -0.25cm
\caption{Left panel:  comparison between the
completeness-corrected STIS LP luminosity functions for the Ursa Minor
dSph (solid line) and the globular cluster M15 (points).  Right panel:   
comparison
between the  STIS-based  I-band luminosity function of the Ursa
Minor dSph (solid line), the observed I-band luminosity function (points), and the
luminosity function corresponding to a power-law mass function with
slope $-1.35$ (dashed line).}
\end{figure}
\vskip -0.25cm

\subsection{Mass Functions}

The luminosity function corresponding to a mass function of slope
$-1.35$ (where the Salpeter slope is $-2.35$) may be calculated using
the Baraffe {\it et al.} (1997) models. This provides a reasonable fit to the data for M92 and for
M15 (Piotto \& Zoccali 1999).  Further, Zoccali {\it et al.} (1999) 
found this mass
function to fit their data for the faint stellar luminosity function of
the bulge of the Milky Way in Baade's window. 
Figure~2  shows our WFPC2 I-814 luminosity function
for Ursa Minor, together with that derived from our transformed STIS LP
data and  the predictions from the Baraffe {\it et al.} models. 
This mass function appears
to be an excellent  description of the underlying IMF  
in the Ursa Minor dSph too. 

\section  {Conclusions}

As described here, the main sequence stellar luminosity function of
the Ursa Minor dSph, and implied IMF, down to  $\sim
0.4$M$_\odot$, is indistinguishable from that of the halo globular
clusters M92 and M15, systems with the same old age and low
metallicity as the stars in the Ursa Minor dSph.  However, the
globular clusters show no evidence for dark matter, while the Ursa
Minor dSph is apparently very dark-matter-dominated. 
Thus the low mass stellar IMF for stars that formed at high redshift
is invariant in going from a low-surface-brightness, 
dark-matter-dominated external galaxy, to a globular cluster within the Milky Way.
This luminosity function, and underlying IMF, is in good agreement  with
those derived for the field stars of the Milky Way bulge  and is 
consistent with 
the field halo and disk (e.g. von Hippel {\it et al.} 
1996), supporting the concept of a universal IMF.

\acknowledgments

RFGW thanks the Center for Particle Astrophysics, UC Berkeley, for
hospitality.  Support for this work was provided by NASA grant
number GO-7419 from STScI, operated by AURA Inc, under NASA
contract NAS5-26555.  We thank J.~Gallagher, N.~Tanvir \&
T.~Smecker-Hane for discussions.

\end{document}